# Integrated Gallium Nitride Nonlinear Photonics

Yanzhen Zheng[1], Changzheng Sun[1,*], Bing Xiong[1], Lai Wang[1], Zhibiao Hao[1], Jian Wang[1], Yanjun Han[1], Hongtao Li[1], Yu Jiadong[1], Yi Luo[1]

[1]Beijing National Research Centre for Information Science and Technology (BNRist), Department of Electronic Engineering, Tsinghua University, Beijing 100084, China
*Corresponding author: czsun@tsinghua.edu.cn

**Gallium nitride (GaN) as a wide-band gap material has been widely used in solid-state lighting. Thanks to its high nonlinearity and high refractive index contrast, GaN-on-insulator (GaNOI) is also a promising platform for nonlinear optical applications. Despite its intriguing optical proprieties, nonlinear applications of GaN have rarely been studied due to the relatively high optical loss of GaN waveguides (typically ~ 2 dB/cm). In this letter, we report GaNOI microresonator with intrinsic quality factor over 2 million, corresponding to an optical loss of 0.26 dB/cm. Parametric oscillation threshold power as low as 8.8 mW is demonstrated, and the experimentally extracted nonlinear index of GaN at telecom wavelengths is estimated to be $n_2 = 1.2 \times 10^{-18}$ m$^2$W$^{-1}$, which is comparable with silicon. Single soliton generation in GaN is implemented by an auxiliary laser pumping scheme, so as to mitigate the high thermorefractive effect in GaN. The large intrinsic nonlinear refractive index, together with its broadband transparency window and high refractive index contrast, make GaNOI a most balanced platform for chip-scale nonlinear applications.**

Gallium nitride (GaN) as a wide-band gap material has been extensively investigated for semiconductor lighting over the past decades. Epitaxial growth of GaN on sapphire (GaNOI) by metal-organic chemical vapor deposition (MOCVD) is quite mature, and GaN thin films with excellent crystalline quality can be routinely obtained. The refractive index of GaN at telecom wavelengths is about 2.3 and the GaNOI platform offers high refractive index contrast (~ 0.6). Thanks to the large band gap of GaN (3.4 eV), GaN exhibits a wide transparency window, ranging from ultra-violet to mid-infrared. In addition, wurtzite GaN is known to be highly resilient against harsh environment, such as high temperature or high optical power. Thanks to its non-centrosymmetric crystal structure, GaN possesses both $\chi^{(2)}$ and $\chi^{(3)}$ nonlinearities, which makes it appealing for electrically tuned chip-based photonic applications as well as second-harmonic generation (SHG). Previous study shows that the nonlinear refractive index of GaN film at telecom wavelengths is ~$10^{-18}$ m$^2$W$^{-1}$, which is an order of magnitude larger than that of conventional platforms [1,2], such as Si$_3$N$_4$ [3,4], AlN [5,6] and LiNbO$_3$ [7,8]. These unique properties make GaNOI platform attractive for compact chip-scale nonlinear photonic applications, such as frequency conversion, supercontinuum and frequency comb generation. However, apart from very limited reports on four-wave mixing (FWM) and frequency conversion [1,2], nonlinear applications of GaNOI have rarely been investigated up to now. The main barrier is the relatively large optical loss of GaN waveguides, typically on the order of 2 dB/cm, corresponding to a quality factor $Q$ ~$10^5$ for microresonators [1,9].

In this letter, we report dissipative Kerr soliton (DKS) generation in high $Q$ factor microring resonators fabricated on GaNOI platform. The intrinsic $Q$ factor is up-to 2.5 million, corresponding to a waveguide loss of 0.17 dB/cm. The threshold for parametric oscillation is as low as 8.8 mW, indicating a nonlinear refractive index of $n_2 = 1.2 \times 10^{-18}$ m$^2$W$^{-1}$.

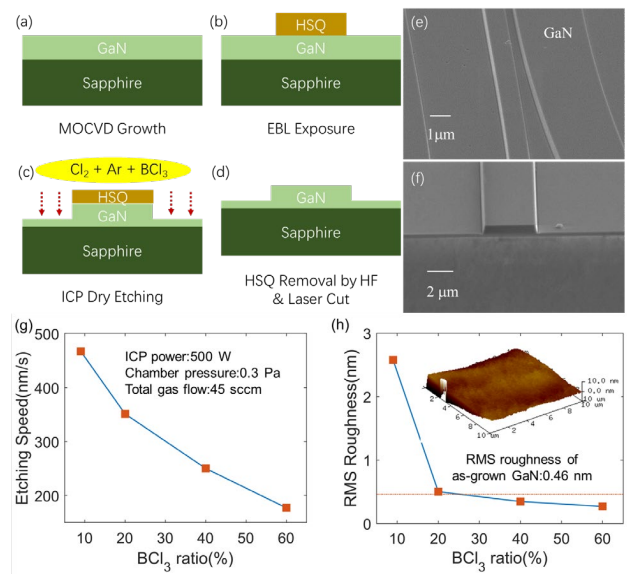

Fig. 1 (a-d) Fabrication process and (e, f) scanning electron microscope (SEM) images of the high $Q$ GaNOI microresonator. The GaN film undergoes electron beam lithography (EBL), inductively coupled plasma (ICP) etching, mask removal and laser cleaving process. (g) Etching rates and (h) surface roughness as a function of BCl$_3$ ratio. The ICP power and the RF bias are set to 500 W and −300 V, while the total gas flow and the chamber pressure are fixed at 45 sccm and 0.3 Pa, respectively. The gas ratio of Cl$_2$/Ar is kept at 3:1. The inset shows the atomic force microscope (AFM) image of surface morphology of GaN sample obtained with optimized etching conditions.

The microring resonator is fabricated on a 1-μm-thick un-doped GaN film epitaxially grown on a $c$-plane sapphire substrate by MOCVD. The GaN film thickness is so chosen as to ensure strong light confinement in the waveguide as well as high crystalline quality. The pattern for microring and associated bus waveguide is formed by electron-beam lithography (EBL) with 600-nm-thick hydrogen silsesquioxane (HSQ) as hard mask. The microring is then dry etched by inductively coupled plasma (ICP) based on Cl$_2$/Ar/BCl$_3$ mixture. The dry etching process is optimized to ensure smooth etched surface as well as vertical sidewalls [9]. It is found that the BCl$_3$ ratio has a crucial impact on etch rate as well as surface morphology. As shown in Fig. 1(g, h), both the etch rate and the surface roughness decrease with increased BCl₃ ratio. However, a BCl₃ ratio greater than 40% would lead to angled sidewalls. Consequently, a BCl₃ percentage of 30% is adopted, which ensures a moderate etching rate of 295 nm/min as well as a root-mean-square (RMS) roughness of 0.50 nm, comparable to that of the as-grown GaN film (0.46 nm). The HSQ layer was removed by 40% HF solution after the ICP etching. To avoid the scattering loss due to voids formed in the top cladding during plasma enhanced chemical vapor deposition (PECVD), an air-clad GaN microring resonator is adopted in this work [10]. Finally, the chip was cleaved by a high power laser to enable efficient optical coupling. The fabrication process as well as the scanning electron microscope (SEM) images are shown in Fig. 1. The SEM images reveal smooth etched surface as well as nearly vertical sidewalls.

Kerr combs based on high $Q$ microresonators have attracted much attention, and of particular interest are microcombs operating in DKS states, which enable a variety of applications such as low phase-noise microwave generation, optical frequency synthesis and photonic radar [11]. DKS generation normally requires microresonators with anomalous dispersion. However, similar to most materials, GaN exhibits normal material dispersion at telecom wavelengths. The microresonator therefore needs to be dispersion-engineered by adjusting the ring waveguide geometry to meet the dispersion requirement. The simulated dispersion profiles for $TE_{00}$ and $TM_{00}$ modes in a 60-μm-radius GaN microring are shown in Fig. 2(c). Broadband anomalous dispersion for both $TE_{00}$ and $TM_{00}$ is secured with 2.25-μm waveguide width and 730-nm etching depth. The mode profiles shown in the inset of Fig. 2(a) indicate strong light confinement, with a mode area of 2.8 and 2.9 μm² for $TE_{00}$ and $TM_{00}$, respectively. A schematic of the GaNOI microring is shown in Fig. 2(a). Pulley coupled bus waveguide with a width of 1.1 μm and a coupling gap of 650 nm is adopted to realize efficient coupling. The width of the bus waveguide is expanded to 3.3 μm at both facets via a 400-μm-long taper section to reduce the fiber-to-chip coupling loss. The measured transmission spectra of the GaNOI microring is shown in Fig. 2(b), which shows a free spectral range (FSR) of 330 and 324 GHz for $TE_{00}$ and $TM_{00}$ modes, respectively. The insertion loss is estimated to be 3.7 dB per facet for both $TE_{00}$ and $TM_{00}$ modes. The measured and fitted resonance curves for $TM_{00}$ mode at 1550 nm are shown in Fig. 2(d). The extracted intrinsic $Q$ factors are 2.5×10⁶ and 1.8×10⁶ for $TE_{00}$ and $TM_{00}$ modes, corresponding to a propagation loss of 0.17 and 0.24 dB/cm, respectively. The intrinsic quality factor of our GaN microresonator is comparable to that commonly attainable with $Si_3N_4$ or $LiNbO_3$ platform.

Parametric oscillation in the GaN microring occurs when continuous wave pump light with sufficient power is tuned into resonance. The $\chi^{(3)}$ nonlinearity of GaN material can be extracted by measuring the parametric oscillation threshold $P_{th}$ (see supplement 1 for more information). The parametric oscillation threshold is estimated to be 8.8 mW, corresponding to a nonlinear refractive index $n_2$ = 1.2×10⁻¹⁸ m²W⁻¹. This value is consistent with previous reported results [1].

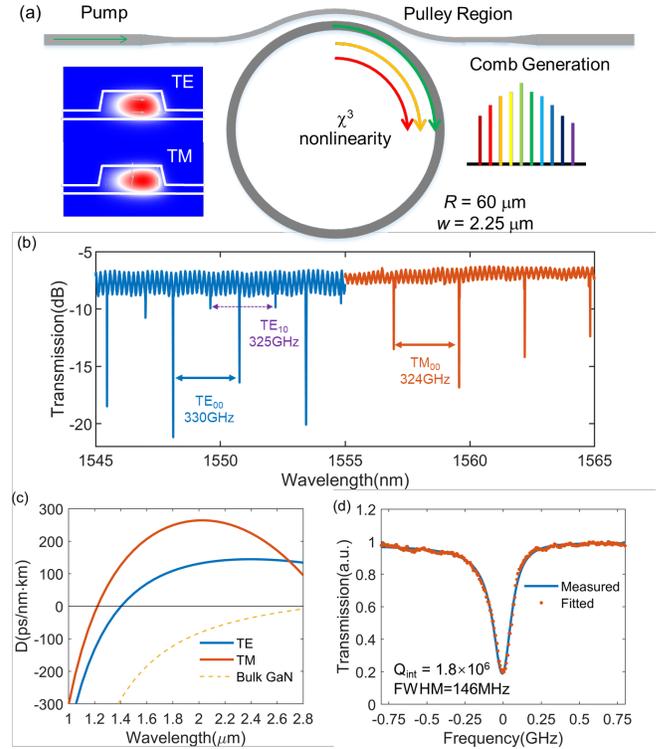

Fig. 2 (a) Top view of the GaN microring resonator. The inset at the left bottom shows the simulated $TE_{00}$ and $TM_{00}$ modes inside the cavity. (b) Transmission spectra of the GaN microring. The FSR of the 60-μm-radius microresonator is 330 GHz and 324 GHz for $TE_{00}$ and $TM_{00}$ modes, respectively. (c) Simulated dispersion profile of the dispersion-engineered microresonator with a waveguide width of 2.25 μm. (d) Measured resonance linewidth and extracted Q-factors for $TM_{00}$ mode. The blue dots and the red line correspond to measured and fitted results, respectively.

Broadband optical frequency comb can be generated with the dispersion-engineered high $Q$ GaN microresonator. As shown in Fig. 3(f), octave-spanning chaotic comb is recorded at high pump power (on chip power ~ 450 mW), ranging from 125 to 250 THz. In addition to Kerr comb lines, peaks arising from stimulated Raman scattering (SRS) can also be identified, corresponding to excitation of $E_2$(high) and $A_1$(TO) phonons in the wurtzite crystal [12].

Such chaotic comb exhibits high noise and corresponds to blue-detuned pump laser with respect to the hot cavity mode. Low noise comb in DKS regime corresponds to red-detuned pump laser with respect to resonance, and requires overcoming transient thermal effects induced by the sharp reduction in intracavity power during transition to soliton states [11]. Such thermal instability is particularly pronounced in materials with high thermorefractive coefficient d$n$/d$T$ (e.g. ~ 3.6×10⁻⁴ K⁻¹ for AlGaAs), thus rendering soliton states thermally inaccessible [13,14]. GaN is reported to exhibit a thermorefractive coefficient ~10⁻⁴ K⁻¹ at telecom wavelengths [15], comparable with that of AlGaAs. Previously, soliton generation in high d$n$/d$T$ material is reported by reducing the thermorefractive coefficient through cryogenic cooling [13], which severely limits its use in many practical applications. Here, we adopt auxiliary laser pumping scheme to mitigate thermal instability [10,16,17] in a GaN microresonator and stably access soliton states at room temperature.

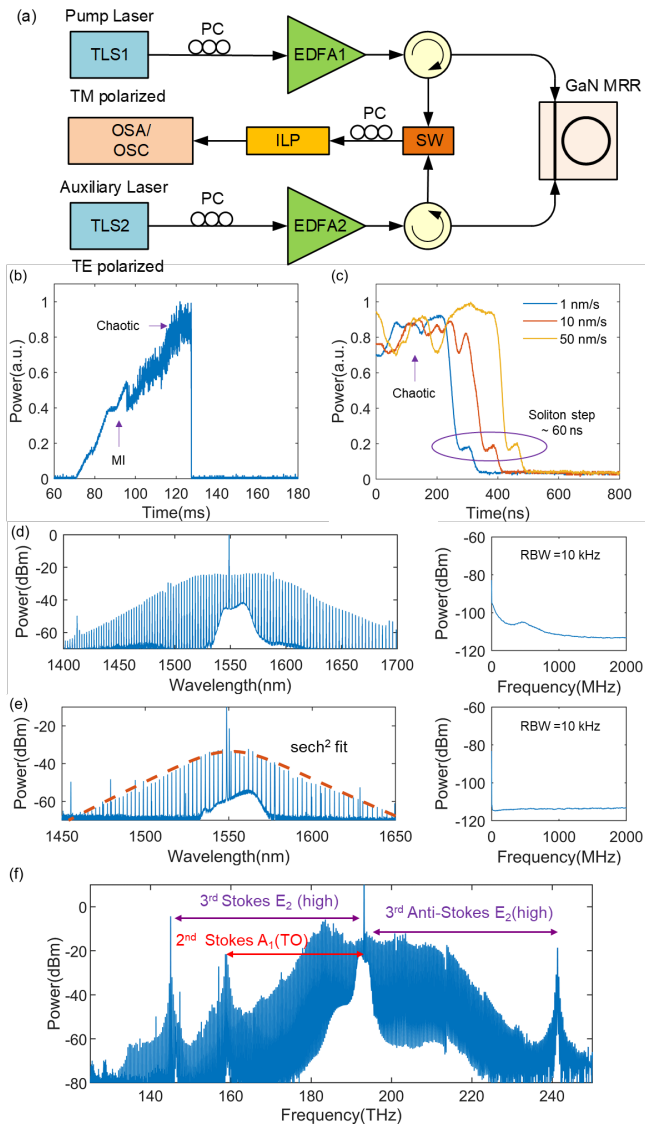

with different sweep speeds are plotted in Fig. 3(c). The soliton-step is extremely short (~60 ns) and remains basically unvaried as the scan speed is increased by 50 fold, indicating strong thermal effects in GaN microresonator during soliton transition [18]. To mitigate the significant thermal effects, light from an auxiliary laser is counter-coupled into the GaN microresonator. When the auxiliary laser is adequately positioned with respect to the resonance, thermal instability induced by intracavity power drop can be compensated, and the total intracavity power remains nearly unvaried as the pump laser is swept to the red-detuned regime, thus enabling stable access to soliton states [10,16,17].

In our experiment, the pump laser is TM-polarized while the auxiliary laser is TE-polarized. Both modes are in nearly critical coupling regime and offer efficient thermal compensation. Adopting orthogonally polarized pump and auxiliary lights allows simple separation of the DKS comb from the reflected auxiliary comb with an in-line polarizer. The on-chip power for both the pump and the auxiliary lasers is estimated to be 136 mW. Soliton state is accessed by fine adjusting the detuning and power of both pump and auxiliary lasers (see supplement 1 for more information). As shown in Fig. 3(e), the single soliton comb features an optical spectrum with a smooth $sech^2$ envelope as well as low RF noise. For comparison, the optical and RF noise spectra for the chaotic state are plotted in Fig. 3(d). It is seen that the chaotic comb exhibits high noise below 2 GHz.

Compared with soliton generation at cryogenic temperature, the auxiliary laser pumping offers simple and stable access to soliton states in high thermorefractive coefficient materials. In addition, the cryogenic cooling scheme is plagued by a limited pump power window over which soliton states are thermally accessible, as the thermorefractive coefficient increases with the pump power [13]. On the other hand, soliton generation in a dual-pumped microresonator is not limited by such pump power window, as long as the thermal effects is effectively mitigated by the auxiliary laser. (see supplement 1 for more information).

To date, a variety of platforms have been developed for on-chip nonlinear applications. Table 1 summarizes several material platforms where soliton generation has been reported at telecom wavelengths. As shown in Table 1, the nonlinear refractive index of GaN is comparable with that Si and about an order of magnitude larger than that of $Si_3N_4$, $LiNbO_3$ or AlN. Meanwhile, the two-photon absorption (TPA) cutoff wavelength of GaN is 729 nm, which is significantly shorter than that of $Al_{0.17}Ga_{0.83}As$, GaP and silicon. Thanks to the large bandgap of GaN, the transparency window of GaNOI ranges from ultra-violet to mid-infrared, and has the potential to enable visible comb generation important for optical clocks [19]. Furthermore, $LiNbO_3$, AlGaAs or GaP based nonlinear optics often involve wafer bonding process to realize effective light confinement, whereas GaN film with excellent crystalline quality can be directly grown on sapphire substrate. Mature MOCVD growth technique guarantees precise film thickness control for dispersion tailoring in microcombs, and the large refractive index contrast ensures strong optical confinement important for nonlinear optics applications. These intrinsic material properties make GaN the most balanced platform for chip-scale nonlinear applications.

In conclusion, we present GaN-on-sapphire as a promising platform for nonlinear photonics. Thanks to the large nonlinear refractive index of GaN, parametric oscillation threshold as low as 8.8 mW is recorded for a 60-μm-radius microring with intrinsic quality factor exceeding $10^6$. Stable access to soliton regime in high-$Q$ GaN microring resonators is demonstrated by employing auxiliary laser pump scheme to mitigate its high thermorefractive effect. Compared with other nonlinear platforms

Fig. 3 (a) Experimental setup for DKS generation in a GaN microresonator. Circulators are employed to separate light entering and exiting the GaN microring resonator (MRR). TLS, tunable laser; EDFA, erbium-doped fiber amplifier; OSA, optical spectrum analyzer; OSC, Oscilloscope; PC, polarization controller; ILP, inline-polarizer; SW, optical switch. (b) Converted comb power trace recorded with a scan speed of 1 nm/s. (c) Observed soliton-step duration for various sweep speeds. Optical spectrum and RF noise spectrum of single comb line for (d) chaotic comb and (e) single-soliton comb. The 3 dB band width of the single soliton is ~ 50 nm with an FSR of 324 GHz. (f) Octave-spanning chaotic Kerr-Raman combs. The pump is TE-polarized with 560 mW on-chip power. The peaks at around 240 and 145 THz is attributed to the third order $E_2$(high) phonons, while the peak at 158 THz is attributed to the second order $A_1$(TO) phonons, with a Raman shift of 575 and 579 $cm^{-1}$, respectively.

The experimental setup for soliton generation is shown in Fig. 3(a). First, the influence of thermal effects on soliton-step duration is investigated. Soliton-steps in the GaNOI microresonator are captured by fast sweeping the pump laser across the resonance. The converted comb power trace recorded with a pump power of 120 mW at a sweep speed of 1 nm/s is shown in Fig. 3(b), while the soliton-steps obtained reported so far, GaNOI is unique in its balanced performance in terms of high intrinsic nonlinearity, high refractive index contrast as well as wide transparency window. In addition, having both $\chi^{(2)}$ and $\chi^{(3)}$ nonlinearity makes GaN attractive for broadband Kerr comb and supercontinuum generation on chip.

Table 1. Properties of nonlinear platforms for chip-scale frequency comb generation operating at telecom wavelengths

| Material | $n$ | $\chi^{(2)}$ (pm V$^{-1}$) | $n_2$ (10$^{-18}$ m$^2$W$^{-1}$) | $\lambda_{TPA}$ (nm) [a] | Mode Area ($\mu$m$^2$) | FSR (GHz) | $Q_{int}$ (×10$^6$) | $P_{th}$ (mW) | Remarks |
|---|---|---|---|---|---|---|---|---|---|
| Si [20] | 3.5 | – | 4 | 2250 | – | – | – | – | – |
| Al$_{0.2}$Ga$_{0.8}$As [21] | 3.3 | – | 26 | 1483 | 0.28 | 1000 | 1.5 | ~0.03 | Bonding |
| Si$_3$N$_4$ [4] | 2 | – | 0.25 | 460 | ~1 | 99 | ~10 | <1 | – |
| AlN [5,6] | 2.1 | 0.43 | 0.23 | 440 | 2.3 | 435 | 0.8 | 25 | MOCVD growth |
| Diamond [22] | 2.4 | – | 0.82 | 450 | 0.81 | 925 | 0.97 | 20 | – |
| GaP [23] | 3.1 | 82 | 11 | 1100 | 0.15 | 250 | 0.2 | 3 | Bonding, no soliton |
| LiNbO$_3$ [7] | 2.2 | 54 | 0.18 | 635 | 1 | 200 | ~4 | 4.2 | Bonding |
| GaN (this work) | 2.3 | –9 [b] | 1.2 | 729 | 2.9 | 324 | 1.8 | 8.8 | MOCVD growth |

[a] $\lambda_{TPA}$ corresponds to two-photon absorption induced cutoff wavelength of the material.
[b] The value of $\chi^{(2)}$ can be found in [24].

**Funding** This work was supported in part by National Key R&D Program of China (2018YFB2201700); National Natural Science Foundation of China (61975093, 61927811, 61822404, 61974080, 61904093, and 61875104); Tsinghua University Initiative Scientific Research Program (20193080036), The China Post-doctoral Science Foundation (2018M640129, 2019T120090).

**Disclosures.** The authors declare no conflicts of interest.